\documentclass[fleqn,usenatbib]{mnras}

\usepackage[T1]{fontenc}

\DeclareRobustCommand{\VAN}[3]{#2}
\let\VANthebibliography\thebibliography
\def\thebibliography{\DeclareRobustCommand{\VAN}[3]{##3}\VANthebibliography}

\usepackage{graphicx}	
\usepackage{amsmath}	
\usepackage{amssymb}	

\def\gtorder{\mathrel{\raise.3ex\hbox{$>$}\mkern-14mu
             \lower0.6ex\hbox{$\sim$}}}
\def\ltorder{\mathrel{\raise.3ex\hbox{$<$}\mkern-14mu
             \lower0.6ex\hbox{$\sim$}}}

\title[Time Delays]{Measuring time delays: II. Using observations of the unresolved flux and astrometry}

\author[Springer \& Ofek]{Ofer M. Springer,$^{1,2}$
Eran O. Ofek$^{1}$\thanks{E-mail: eran.ofek@weizmann.ac.il}
\\
$^{1}$Department of Particle Physics and Astrophysics, Weizmann Institute of Science, 76100 Rehovot, Israel.\\
$^{2}$Benin School of Computer Science and Engineering, The Hebrew University of Jerusalem, 9190416 Jerusalem, Israel.}

\pubyear{2015}

\begin{document}

\label{firstpage}
\pagerange{\pageref{firstpage}--\pageref{lastpage}}
\maketitle

\begin{abstract}

Lensed quasars and supernovae can be used to
study galaxies' gravitational potential and measure cosmological parameters.
The typical image separation of objects lensed by galaxies is of the order
of $0.5''$. Therefore, finding the ones with small
separations, and measuring their time-delays using
ground-based observations is challenging.
We suggest a new method to identify lensed quasars and simultaneously measure their time-delays, using seeing-limited synoptic observations
in which the lensed quasar images and the lensing galaxy are unresolved.
We show that using the light curve of the combined flux, and the astrometric measurements of the center-of-light
position of the lensed images,
the lensed nature of a quasar can be identified,
and its time-delay can be measured.
We provide the analytic formalism to do so, taking into account
the measurement errors and the
fact that the power spectra of quasar light curves is red
(i.e., the light curve is highly correlated).
We demonstrate our method on simulated data,
while its implementation to real data will be presented in future papers.
Our simulations suggest that, under reasonable assumptions, the new method can detect unresolved lensed quasars and measure their time delays,
even when the image separation is below $0.1''$,
or the flux ratio between the faintest
and brightest images is as low as 0.03.
Python and MATLAB implementations are provided.
In a companion paper, we present a method for measuring the time delay
using the combined flux observations.
Although the flux-only method is less powerful, it may be useful in cases
in which the astrometric information is not relevant (e.g., reverberation mapping).

\end{abstract}


\section{Introduction}
\label{sec:intro}

Lensed quasars and supernovae allow us to estimate the mass of lensing galaxies (e.g., \citealt{Maoz+Rix1993_LensedQSO_Stat}, 
\citealt{Treu+2006_LensedGalSurveyIII_GalProfile});
measure the mass evolution of galaxies (e.g., \citealt{Ofek+2003_LensedQSOz_GalEvolution});
probe the size of the accretion disk around massive black holes (e.g., \citealt{Kochanek2004_QSO_AccDisk_Microlensing}, \citealt{Morgan+2010_QSO_AccreationDiskSize_MassBH}, \citealt{Chan+2020_QSO_AccreationDiskSize_QSO_Microlensing});
measure the stellar mass of galaxies (e.g., \citealt{Poindexter+Kochanek2010Q2237_transverse_velocity_microlensing});
study the host galaxies of quasars;
estimate cosmological parameters;
and measure the Hubble constant via the time delays between the light curves
of the lensed images (e.g., \citealt{Refsdal1964_HubbleConstant_FromLensingTimeDelay},
\citealt{Saha+2006_HubbleTime_LensedQSO_TimeDelay}, \citealt{Oguri2007_LensTimeDelaySample_H0}, \citealt{Bonvin+2017_HE0435_TimeDelay_H0}, \citealt{Birrer+Treu2020_TimeDelay_H0_Strategies};
see however, \citealt{Kochanek2002_WhatTimeDelayMeasure}, \citealt{Kochanek2020_LensedQSO_TimeDElay_MeasuringH0_OverConstr};
\citealt{Blum+2020_LensedQuasars_H0_Core};
\citealt{Morstell+2020_LensModelingOfLenseIaSN_PTF16geu}).

Finding lensed objects is challenging. Their typical image separations
is of the order of $0.5''$, just below the seeing disk induced by Earth's atmosphere.
Furthermore, measuring their time delays requires partially resolving the images,
which is difficult.
There are over 100 known lensed quasars, but for only $\sim30$ of them there are measured time delays.
So far, only a few cases of lensed supernovae have been found
(e.g., \citealt{Quimby2014_LensedIaSN}; \citealt{Goobar+2017_PTF16geu_LensedIaSN}),
but more are expected to be found using future sky surveys data.

Several methods have been used to detect lensed quasars.
Among them are diffraction-limited space-based observations (e.g., \citealt{Maoz+1992_HST_LensedQSO}; \citealt{Lemon+2019_22newLensedQuasarGAIA}),
observations from good-seeing sites (e.g., \citealt{Morgan+2004_WFI2026_WFI2033_LensedQSO}),
multi-band selection of slightly extended sources
(e.g., \citealt{Ofek+2007_LensedQSO_J1313+5151}, \citealt{Jackson+2012_NewLensedQiasars}),
radio observations (e.g., \citealt{Myers+1995_CLASS_1608_LensedQSO}), and more.
\cite{Kochanek2006_FindingLensedQSO_Variability} suggested
that it may be possible to find lensed quasars based on the fact that they
are variable and slightly extended sources.
For finding unresolved lensed Type Ia supernovae (SNe),
\cite{Goldstein+2017_HowToFindLensedIaSN}
suggested to look for transients
in elliptical galaxies.
For such galaxies, the photometric redshift is reliable;
in this case, we can look for a Type Ia SN light curve
that is more luminous than typical Type Ia SN at the host distance.
This method is limited to a fraction of Type Ia SNe that take place in elliptical galaxies.
Additional options include fitting the known (or partially known)
light curve shape with time delayed copies of itself
to the observed light curve (e.g., \citealt{Dhawan+2019_TimeDelay_Lensed_SNIa_PTF};
\citealt{Bag+2020_MethodTimeDelay_LensedSN}).

Measuring the time delays between the images of lensed quasars
and SNe
is typically more challenging than finding the lenses.
With the exceptions of large separation lenses, which are
completely resolved (e.g., \citealt{Inada+2003_SDSS1004_LensedQSO}; \citealt{Fohlmeister+2007_SDSS1004+4112_TimeDelay}), such measurements
require excellent seeing observations (e.g., \citealt{Burud+2002_HE2149_LensedQSO_TimeDelay}).
Furthermore, given that the typical time delays are of the order of months, the correlated structure of quasar light curves makes
it even more difficult to measure the time delays.
For example, \cite{Ofek+2003_HE1104-1805_TimeDelay}
demonstrated that ignoring the correlations in quasar
light curves results in significant underestimation of the time-delay
uncertainties.
Alternatively,
\cite{Geiger+Schneider1996_LensedQuasarTimeDelay_AutoCorrelationFunction}, \cite{Pindor2005_LensedQuasarTimeDElay_AutoCorrelationFunction}
and \cite{Shu+2020_LensedQuasarTimeDElay_AutoCorrelationFunction} suggested using the auto-correlation (or related) function of the combined light curve of lensed quasars to measure their time delay.
In Paper~I (Springer \& Ofek)
we derive the likelihood function of the combined-flux of several time delayed light curves, given the free parameters (e.g., time delay, flux ratio).
Our derivation takes into account the fact that quasar light curves are generated from a red-noise process.
The likelihood function found in Paper~I is different than the auto-correlation function.

In this paper, our main goal is to derive a technique
that will enable us to identify lensed variable sources (i.e., quasars and SNe)
among non-lensed objects
and simultaneously measure their time-delays
from a set of multi-epoch observations
in which the lensed object images are unresolved.
We demonstrate that the information about the individual light curves
of the lensed object images is contained in the
combined flux and center-of-light position, as a function of time.
When the image-separation is smaller than a fraction of the seeing-disk,
unless one has an accurate and detailed knowledge of the PSF,
it is difficult to achieve similar results using PSF-fitting or de-convolution.
In order to derive such a method, we first build a statistical model that
describes the observations.
Next, we derive the likelihood of the observations
(total flux and centroid position)
given the model and the free parameters
(e.g., the time-delay and flux ratio).
An important ingredient of our modeling is that it
accounts for the fact
that quasar light curves are highly correlated and their power spectrum
is red (e.g., \citealt{Markowitz+2003_Quasars_AGN_Xray_powerspectra}, \citealt{Mushotzky+2011_AGN_PowerSpectra_KeplerLC}, \citealt{Smith+2018_AGN_KeplerLC_PowerSpectrum}).
Neglecting this fact will bias the results and will result in underestimation
of the uncertainties.

The likelihood function of the combined flux given the free parameters, derived in Paper~I,
is used in the current work.
For lensed quasars, the flux-only technique is not as powerful as the flux plus astrometry method
as it uses only part of the available information.
However, the flux-only method
is relevant for systems in which the astrometry
does~not provide any additional information
(e.g., reverberation mapping).

The structure of this paper is as follows.
In \S\ref{sec:description} we qualitatively describe our algorithm, while in \S\ref{sec:method} we derive the likelihood functions required in order to apply our method.
In \S\ref{sec:Implimintation} we
discuss the implementation details,
while our code is described in \S\ref{sec:Code}.
In \S\ref{sec:simulations} we test our new method on simulated data.
Finally, in \S\ref{sec:disc} we conclude and discuss the method's caveats, and some of its possible extensions.

\section{Schematic description of the method}
\label{sec:description}

We would like to find lensed quasars and measure their time delays using observations of the combined flux and center-of-light position (astrometry).
For simplicity and clarity, all our discussions are for the case of a lensed quasar with two images.
In Appendix~\ref{sec:multi_image} we extend this method to lensed quasars with an arbitrary number of images.

In the case that both the total flux and center of light position are available, it is clear that
a solution exist.
Specifically, the center of light position provides information on the flux ratio between
the images, and along with the total fluxes, the light curves of the individual images
can be reconstructed, and therefore the time-delay can be measured.
A caveat of this approach is that the light curves of the
individual images
from the time of the first observation to 
the time of the first observation plus the maximum time delay,
is not well reconstructed.
This is because for such early (or late) observations, the total flux depends on
the individual light curves prior to the first observation (or after the last observation).
Therefore, in order for this method to work, it is required that the total time span
of the observations will be considerably larger than the time delay.
As shown in Paper~I, the problem can be solved using only the combined flux.
By adding more information,
we expect that the solution will improve.

An important question is 
if it is feasible to
detect the center-of-light motion of lensed quasars using ground-based observations.
The most important contribution to the astrometric
noise in such observations is the
scintillations in Earth's atmosphere (e.g, \citealt{Lindegren1980_AtmosphericScintilation_GroundBasedAstrometry}; \citealt{Shao+Colavita1992_AtmosphericScintilation_GroundBasedAstrometry}).
On scales of about ten arcminutes, which are relevant for
wide-field synoptic surveys and stellar surface density at high Galactic latitudes, the contribution
of the scintillation noise is of the order of
10\,mas for telescopes of about 1\,m and exposure times
of about 1\,min (e.g., \citealt{Ofek2019_Astrometry_Code}).
Since quasars' photometric variability is of the order of
$\gtorder10$\%, and the typical image separation
of lensed quasars is of the order of $0.5''$,
we expect the center-of-light motion of lensed
quasars to be above the expected
astrometric noise level
(even in a single epoch).

Another question is why use only the zeroth moment (total flux)
and center-of-light position (first moment),
and not the full shape information of the blended lensed quasar (higher moments)?
In other words, can we reconstruct the light curve of the individual
lensed images by either forward modeling (i.e., PSF fitting; e.g., \citealt{Ofek+2003_HE1104-1805_TimeDelay}),
or deconvolution of the blended lensed images (e.g., \citealt{Burud+2000_B1600_LensedQSO_TimeDelay})?
These two methods are equivalent (at least at the limit of high $S/N$),
and both methods were used successfully in the past.
However, when the lensed quasar image separation is considerably smaller than the
point spread function full-width at half maximum (FWHM),
PSF fitting and deconvolution are very sensitive to errors in the PSF.
For example, when the images' separation are of the order of the FWHM of unresolved sources,
the errors in photometry are roughly linear with the errors in the PSF.
Given the non-perfect knowledge of the PSF, these methods will break down
at some point.
Therefore, using only the zeroth and first moment considerably simplifies
the situation and removes our dependence on very good knowledge of the PSF.

\section{Method formalism}
\label{sec:method}

Given a quasar light curve and astrometric time series,
our goal is to derive a formalism to identify lensed quasars
and simultaneously measure their time delay when the quasar light curve
has a red power spectrum.
The observables are both the total flux from the lensed images and lensing galaxy, as well as their combined
center-of-light position as a function of time.
A lensed supernova can be regarded as a 
variant in which we have an exact model 
for the light curve in the time domain.

In order to achieve these goals we are interested in the expressions of the likelihood
of the observations given the model and its free parameters,
where the free parameters may include the
time delay and flux ratio between the images, the lensed image's positions, and the lensing galaxy's flux and position.
This likelihood can be used to measure the free parameters,
and to perform an hypothesis testing between
the null hypothesis that the quasar is not lensed
and the alternative hypothesis that the quasar is lensed
(i.e., using the likelihood ratio test
\citealt{Neyman+Pearson1933_HypothesisTesing}).

In the following subsections, we present the derivation of these likelihoods.
For some formulae, the full derivation is available in
the appendices.
In \S\ref{sec:model}, we present the statistical model for the
flux and center-of-light.
To demonstrate the idea at the base of our method
and to simplify the presentation,
in \S\ref{sec:noiseless}, we treat the case of flux and center-of-light
position without noise and under the assumption that
we do~not have a statistical model for the light curve.
In \S\ref{sec:stat_flux}, we provide
the likelihood of the total flux given the model (which was derived in Paper~I),
while in \S\ref{sec:stat_centroid_given_flux}, we derive
the likelihood of the position
given the flux and the model.
In \S\ref{sec:full_model}, we combine the results from the previous subsections
to write the likelihood of the observations given the free parameters.
Finally, in \S\ref{sec:2D}, we discuss the method's extension to two dimensions.

\subsection{Flux and centroid formation model}
\label{sec:model}

We assume that the only available observables are the time-dependent total observed flux and total observed centroid.
For simplicity, here we discuss only the two-lensed image case,
with the extension to multiple images presented in 
Appendix~\ref{sec:multi_image}.
Following Paper~I, the model for the total observed flux is given by:
\begin{eqnarray}
    F(t) &=& \phi(t) + \epsilon_{F}(t) \nonumber \\
         &=&\alpha_0 + \alpha_1 f(t) + \alpha_2 f(t+\tau) + \epsilon_{F}(t),
    \label{eq:F}
\end{eqnarray}
where $\phi(t)$ is the original combined flux (without observational noise), $\alpha_0$ is the flux of the non-variable lensing galaxy,
$f(t)$ is the light curve of the unlensed quasar
as a function of the time $t$, $\tau$ is the time delay between images 1 and 2,
$\alpha_{i}$ is the mean flux of the $i$-th image,
and $\epsilon_{F}(t)$ is the noise in the combined flux
measurement. We assume that the observations are background-noise dominated
and that  $\epsilon_{F}(t)$ is an independent and 
identically distributed (i.i.d.) random Gaussian vector with a per-component variance of $\sigma_{F}^2$. $\alpha_{2}/\alpha_{1}$ is the mean flux ratio between the two images.
The requirement for $\epsilon_F$ to be i.i.d can be relaxed,
in which case, we need to use the full covariance matrix
(see also Paper~I).
We can also write Equation~\ref{eq:F} in the frequency domain
\begin{eqnarray}
    \widehat{F}(\omega) \equiv \widehat{\phi} + \widehat{\epsilon}_F(\omega) \notag \\
    = \alpha_{0}\delta(\omega) + (\alpha_1+\alpha_2 e^{i\omega\tau})\widehat{f}(\omega) + \widehat{\epsilon}_F(\omega),
    \label{eq:F_of_omega}
\end{eqnarray}
where $\delta(\omega)$ is the Dirac delta function,
and the hat sign and $\mathcal{F}$ operator represent the Fourier Transform, defined as
\begin{equation}
    \mathcal{F}[f(t)]\equiv\widehat{f}(w) = \int_{-\infty}^\infty f(t)e^{i\omega t} dt.
    \label{eq:FourierTransform}
\end{equation}

We note that in practice, even if we set the measurement error to zero, the light curves of the various images (properly normalized by $\alpha_i$) may not be identical.
A leading reason for this is variability due to microlensing by individual stars in the lensing galaxy (e.g.,
\citealt{Wambsganss+2000_LensedQuasar_Q0957+561_Microlensing}; \citealt{Wambsganss2001_LensedQuasars_Microlensing}).
To simplify the
analysis, we absorb all these variations into the Gaussian noise term $\epsilon_{F}$ (see Paper~I for additional discussions).

We can express the center-of-light position of the blended image as the following weighted average
\begin{eqnarray}
    \vec{x}(t) &=& \vec{\chi}(t) + \vec{\epsilon}_x(t) \nonumber \\
           &=& \frac{\alpha_0\vec{x}_0 + \alpha_1 \vec{x}_1 f(t) + \alpha_2 \vec{x}_2 f(t+\tau)}{\alpha_0 + \alpha_1 f(t) + \alpha_2 f(t+\tau)}+ \vec{\epsilon}_x(t),
    \label{eq:x}
\end{eqnarray}
where $\vec{\chi}(t)$ is the original centroid position (without observational noise),
$\vec{x}_{0}$ represents the two-component sky position of the lensing
galaxy, and $\vec{x}_{i}$ is the sky position of the $i$-th image.
The centroid measurement noise is denoted by $\vec{\epsilon}_{x}(t)$, which we model as a zero mean two-component i.i.d. Gaussian-noise
vector with a per component variance of $\sigma_x^2$.
In reality, there may be some correlations between
$\epsilon_{F}(t)$ and
$\vec{\epsilon}_{x}(t)$.
For example, in the Poisson noise regime
a larger relative errors in the flux
will induce a larger astrometric errors.
However, in practice, for sources
with a signal-to-noise ratios above $\approx30$, the dominant
astrometric error in ground-based
astrometry is scintillation noise,
which is independent of flux errors
(e.g., \citealt{Ofek2019_Astrometry_Code}).
In any case, the scintillation noise 
tends to decorrelate the position and flux errors.

For brevity of notation, in the following derivations we will write the equations involving the two-component sky positions as scalar equations, and these will be valid for each component separately.
An extension to 2-D is discussed in \S\ref{sec:2D}.

In principle, using equations~\ref{eq:F} and \ref{eq:x}, when the noise terms are negligible,
one can immediately isolate $f(t)$ and solve for the time delay. This exercise is demonstrated in \S\ref{sec:noiseless}.
However,
we would like to use the prior knowledge
that quasar light curves have a roughly known power-law power spectrum.

The particular light curve of the observed quasar's first image $f(t)$ is {\it a priori} unknown. We, therefore, model it statistically.
Following Paper~I, we assume that in the frequency domain, quasar light curves have the following Gaussian distribution:
\begin{equation}
    \widehat{f}(\omega) \sim N(0, \sigma_{\widehat{f}}^2(\omega)),
    \label{eq:f_FT_normal}
\end{equation}
where at each angular frequency $\omega$, $\widehat{f}(\omega)$ is a complex number with independent real and imaginary parts, each with a zero mean and a variance of $\sigma_{\widehat{f}}^2(\omega)/2$.
The tilde sign ($\sim$) denotes the distribution of a random variable.
We assume that for frequencies $|\omega| > 0$, $\widehat{f}(\omega)$ is also statistically independent at different frequencies and has a zero mean, $\operatorname{E}\left[\widehat{f}(\omega)\right] = 0$, and that the variance function has the following power law shape:
\begin{equation}
    \sigma_{\widehat{f}}^2(\omega) = \operatorname{E}\left[\widehat{f}(\omega)\overline{\widehat{f}}(\omega)\right] =
    \operatorname{E}\left[| \widehat{f}(\omega) |^{2}\right] = 
    |\omega|^{-\gamma}.
\end{equation}
Here, the bar sign above the hat symbol indicates a complex conjugation after the Fourier Transform,
and $\gamma$ is the power-law index of the power spectrum.
Observations suggest that quasars have $\gamma\approx1.5$--$3.5$
(e.g.,
\citealt{Edelson+Nandra1999_XrayPowerSpectrum_AGN_NGC3516};
\citealt{Mushotzky+2011_AGN_PowerSpectra_KeplerLC};
\citealt{Edelson+2014_Kepler_AGN_Zw229-15_5day_timescale};
\citealt{Kasliwal+2015_IsAGNLCConsistentWithDampedRandomWalk_AGN_PowerSpectrum},;
\citealt{Smith+2018_AGN_KeplerLC_PowerSpectrum}).
We refer the reader to Paper~I for additional discussion.
Note that we use $\sigma_{\widehat{f}}^2$ to denote the
power spectrum of $f$, while we use the notation $\widehat{\sigma}_{f}^{2}$ to denote the
variance of $\widehat{\epsilon}_{f}$.

\subsection{A simplistic example -- Constraining the centroid using the flux in the noiseless case}
\label{sec:noiseless}

To demonstrate that the combination of flux and centroid observations encode the parameters of interest, we derive the following in the case where there is no observational noise,
and the statistical distribution of $f(t)$ is unknown.

\subsubsection{Frequency representation}
\label{sec:noiseless_freq}

We start by defining the {\it non-normalized centeroid} $G(t) \equiv x(t)F(t)$.
Taking the Fourier transform, we find that
\begin{equation}\label{eq:G_of_omega}
    \widehat{G}(\omega) \equiv \mathcal{F}[x(t)F(t)] = (\alpha_1 x_{1}+\alpha_2 x_{2} e^{i\omega\tau})\widehat{f}(\omega),
\end{equation}
when the sky coordinate system is such that the lensing galaxy centroid is $\vec{x}_0 = 0$. By eliminating $\widehat{f}(\omega)$ between equations \ref{eq:F_of_omega} and \ref{eq:G_of_omega}, we find that
\begin{equation}\label{eq:diagnostic}
    \frac{\widehat{G}(\omega)}{\widehat{F}(\omega)} = \frac{\alpha_1 x_{1}+\alpha_2 x_{2} e^{i\omega\tau}}{\alpha_1+\alpha_2 e^{i\omega \tau}}\equiv \widehat{A}(\omega; \tau, \alpha_{i}, x_{i}).
\end{equation}
We see that the left-hand side depends only on the observables, while the right-hand side has a frequency dependence that is only a function of the free parameters $\tau$, $\alpha_i$ and $x_i$. This, therefore, provides a way to find the model parameters by fitting the parametric function $\widehat{A}(\omega)$ to the observations.
For reasons discussed in Paper~I (\S{3.6}), such a direct $\chi^{2}$ fitting
is an incorrect statistical solution to the problem (see below for the proper likelihood function).

\subsubsection{Temporal representation}
\label{sec:noissless_temporal}

By eliminating $f(t+\tau)$ between equations \ref{eq:F} and \ref{eq:x}, we may express the reconstructed quasar flux
\begin{equation}
f_{\rm rec}(t) = \frac{[F(t) - \epsilon_{F}(t)] [x(t) - x_{2} - \epsilon_{x}(t)] + \alpha_{0}[x_{2}-x_{0}]  }{\alpha_{1}(x_{1} - x_{2}) }.
\label{eq:TemporalRepresentation}
\end{equation}
In this expression, positions $x(t)$, $x_{0}$, $x_{1}$, and $x_{2}$ are the positions
projected on the line connecting the two images.
This shows that the reconstruction of $f_{\rm rec}(t)$ does~not requires knowing the time delay $\tau$.
Substituting the reconstructed $f_{\rm rec}(t)$ into $f(t)$ of Equation~\ref{eq:F} we get the reconstructed combined light curve:
\begin{equation}
     F_{\rm rec}(t) =
         \alpha_0 + \alpha_1 f_{\rm rec}(t) + \alpha_2 f_{\rm rec}(t+\tau).
    \label{eq:F}
\end{equation}
Next, we can compare the
reconstructed $F_{\rm rec}(t)$ with the observed $F(t)$,
and fit the free parameters.
For the case of lensed supernova, for which we have a good
model for the light curve shape, this is
the recommended approach.

\subsection{The likelihood of the flux given the model}
\label{sec:stat_flux}

In order to calculate the likelihood of the flux and astrometry given the model,
we first provide the likelihood of the flux given the model.
This likelihood was derived in Paper~I.
Here we briefly repeat the main results and nomenclature.
As discussed in Paper~I, for $\omega\neq0$, we can write:
\begin{eqnarray}
    \widehat{F}(\omega) &=& (\alpha_1+\alpha_2 e^{i\omega\tau})\widehat{f}(\omega) + \widehat{\epsilon}_F(\omega)\\
    & \equiv &  \widehat{\phi}(\omega) + \widehat{\epsilon}_F(\omega).
    \label{eq:Phi_F_of_omega}
\end{eqnarray}
At $\omega=0$, there is an additional additive term of $\alpha_0\delta(\omega)$.
Removing this term is equivalent to subtracting a constant from the data.

The resulting expectation value of the noiseless and noisy total flux is 0.
Following Paper~I,
the expectation of the power spectrum of the light curve
or the variance of the
noisy total flux is
\begin{eqnarray}
    \nonumber
    \Sigma_{F}(\omega) &\equiv& \operatorname{E}\left[\widehat{F}(\omega)\overline{\widehat{F}}(\omega)\right] \\ \nonumber
    &=& \frac{\alpha_1^2+\alpha_2^2+2\alpha_1\alpha_2 \cos(\omega\tau)}{|\omega|^\gamma} +  \widehat{\sigma}_F^2 \\
    &\equiv& \Sigma_{\phi}(\omega) + \widehat{\sigma}_F^2.
    \label{eq:Sigma_F_phi}
\end{eqnarray}
Note that we use the small $\Sigma$ symbol here and in the rest of the text to denote covariance functions and covariance matrices,
while $\Sigma(\omega)$ can be regarded as the diagonal of the covariance matrix.
Finally, given a particular observation $F(t)$ and its frequency representation
$\widehat{F}(\omega)$, the log-probability of observing $F(t)$ given the model parameters is
\begin{eqnarray}
\label{eq:log_P_flux}
    \ln P(\widehat{F} | \tau, \alpha_{i}) = -\frac{1}{2}\ln(\det[2\pi\Sigma_{F}]) - 
    \sum_{\omega}\frac{|\widehat{F}(\omega)|^2}{2\Sigma_F(\omega)}.
    \label{eq:logP_F_par}
\end{eqnarray}
Here, $|\widehat{F}(\omega)|^{2}$ is the power spectrum vector sampled at particular frequencies $\omega$ and $\Sigma_{F}$ is a diagonal covariance matrix (e.g., Equation~\ref{eq:Sigma_F_phi}).
The det operator is the determinant.
In order to avoid having to multiply small numbers, the $\ln{\det}$ should be calculated using the sum of the natural logarithm over the matrix diagonal (see Paper~I). For non-diagonal covariance matrices see Appendix~C in Paper~I.
Equation~\ref{eq:logP_F_par} results from the expression for the log-probability of a multivariate normal distribution (see Appendix \ref{sec:cmvn}).
A graphic representation of Equation~\ref{eq:Sigma_F_phi} is presented in Paper~I.

As discussed in Paper~I,
correlations between frequencies can~not be avoided.
Therefore, it is better to
rewrite this likelihood in the time domain using the full covariance matrix:
\begin{eqnarray}
    \ln P(F | \tau, \alpha_i, \gamma) = -\frac{1}{2}\ln \det[2\pi\Sigma_{T}] \notag \\ 
    - \frac{1}{2} (F(t)-\mu_{\rm F})^{T} \Sigma_{\rm T}^{-1} (F(t)-\mu_{\rm F}).
    \label{eq:LL_TD}
\end{eqnarray}
Here, upper-script $T$ denotes the transpose operation, $\mu_{\rm F}$ is the mean of $F(t)$, and $\Sigma_{\rm T}$ is the covariance matrix between all the pairs in $F(t)$ in the time domain.
Using the Wiener–Khinchin theorem, we can write the elements of the covariance matrix:
\begin{equation}
    \Sigma_{\rm T}(t_{i},t_{j})
    = \int_{-\infty}^{+\infty}{\Sigma_{\rm F}(\omega)e^{-i\omega(t_{i}-t_{j})} d\omega}.
\label{eq:SigmaT_t}
\end{equation}
The integral in Equation~\ref{eq:SigmaT_t} diverges for any value of $\gamma$.
In Paper~I and its appendices, we present two methods to deal with this diverging integral.

\subsection{Statistical model for a centroid given flux}
\label{sec:stat_centroid_given_flux}

Next, we want to derive $P(x|F, \tau, \alpha_1, \alpha_2)$ --- the probability of observing the set of centroids $x(t)$, given that we have already observed a set of total flux measurements $F(t)$. This probability can also be expressed using the previously defined $G(t) \equiv x(t)F(t)$
\begin{eqnarray}
\label{eq:P_x_given_F}
\frac{1}{Z} P(x|F, \tau, \alpha_1, \alpha_2) &=& P(G|F, \tau, \alpha_1, \alpha_2) \nonumber \\
&=& P(\widehat{G}|\widehat{F}, \tau, \alpha_1, \alpha_2),
\end{eqnarray}
where we have linearly transformed the random vector $x(t)$ and, therefore have scaled the probability density by $Z\equiv \Pi_t |F(t)|$. We note that the scalar factor $Z$ depends only on the observations and not on the model parameters. Using equations \ref{eq:F}, \ref{eq:x} and \ref{eq:G_of_omega}, we express $G(t)$ using the noiseless centroid and flux, $\chi (t)$ and $\phi(t)$:
\begin{equation}
G(t) \equiv x(t)F(t) = (\chi (t) + \epsilon_x(t))(\phi (t) + \epsilon_F(t)).
\end{equation}
We now expand the right-hand side, Fourier transform and drop terms that are second order in the noise terms:
\begin{eqnarray}
\label{eq:G_approx1}
\widehat{G}(\omega) \equiv \mathcal{F}[G] &\simeq& \mathcal{F}[\chi\phi] + \mathcal{F}[\chi\epsilon_F] + \mathcal{F}[\phi\epsilon_x].
\end{eqnarray}
Using the definitions of $\phi(t)$ and $\chi(t)$ (Equations \ref{eq:F} and \ref{eq:x}) in the following first equality and the definitions of  $\widehat{\phi}(\omega)$ and $\widehat{A}$ (Equations \ref{eq:F_of_omega} and \ref{eq:diagnostic}) in the following third equality, we find that
\begin{eqnarray}
    \mathcal{F}[\chi\phi] &=& \mathcal{F}\left[ \alpha_0 x_0 + \alpha_1 x_1 f(t) + \alpha_2 x_2 f(t+\tau) \right] \nonumber \\
    &=& \alpha_0 x_0 \delta(\omega) + \left(\alpha_1 x_1 + \alpha_2 x_2 e^{i\omega\tau}\right)\widehat{f}(\omega) \nonumber \\
    &=& \alpha_0 x_0\delta(\omega) + \widehat{A}(\widehat{\phi}-\alpha_0\delta(\omega)),
\end{eqnarray}
which simplifies to $\mathcal{F}[\chi\phi] = \widehat{A}\widehat{\phi}$ at non-zero frequencies. Substituting this result into Equation~\ref{eq:G_approx1} and expanding the other two terms using the convolution theorem, we find (for non-zero frequencies) that
\begin{eqnarray}
\label{eq:G_approx2}
    \widehat{G}(\omega) 
    &\simeq& \widehat{A}\widehat{\phi} + \widehat{\chi}\ast\widehat{\epsilon}_F + \widehat{\phi}\ast\widehat{\epsilon}_x\nonumber \\
    &\simeq& \widehat{A}(\widehat{F}-\widehat{\epsilon}_F) + \widehat{F}\ast\widehat{\epsilon}_x + \widehat{\chi}\ast\widehat{\epsilon}_F \nonumber \\
    &=& \widehat{A}\widehat{F} + \widehat{F}\ast\widehat{\epsilon}_x + \widehat{\chi}\ast\widehat{\epsilon}_F - \widehat{A}\widehat{\epsilon}_F.
\end{eqnarray}
Here, the $\ast$ sign denotes convolution, and we have again dropped second order noise terms on the second line.

We pause to interpret this result. First, we note that this expression for $\widehat{G}(\omega)$ generalizes the result derived in \S\ref{sec:noiseless_freq} to the case where the observational noise is accounted for. Second, this expression shows that we have a simple model for the distribution of the centroid observations once we have observed the combined flux. 

In Appendix \ref{sec:distrib_G_given_F}, we show that $P(\widehat{G}|\widehat{F}, \tau, \alpha_1, \alpha_2)$ is a complex multivariate normal distribution (defined in appendix \ref{sec:cmvn}) with the following mean and covariance:
\begin{eqnarray}
\label{eq:mu_G_given_F}
\mu_{\widehat{G}|\widehat{F}} &=& \widehat{A}\widehat{F} + (\widehat{X}-\widehat{\mathbb{A}})\mu_{\widehat{\epsilon}_{F}|\widehat{F}},\\
\label{eq:Gamma_G_given_F}
\Gamma_{\widehat{G}|\widehat{F}} &=& \widehat{\sigma}_x^2\widehat{F}\widehat{F}^\dagger +
(\widehat{X} - \widehat{\mathbb{A}}) \Gamma_{\widehat{\epsilon}_F|\widehat{F}}
(\widehat{X} - \widehat{\mathbb{A}})^\dagger.
\end{eqnarray}
Here, $\widehat{\mathbb{A}}$ is a diagonal matrix, having the values of $\widehat{A}$ on the diagonal,
$F$ is the vector of flux observations $F(t)$ sampled at a set of discrete times, $X\equiv\mathcal{F}^{-1}\left[\widehat{A}\widehat{F}\right]/F$ is a diagonal matrix with the right-hand side  appearing along the diagonal, and the $\dagger$-sign denotes the transpose complex-conjugate of a matrix (or vector).

Denoting by $X_{jj}$ the diagonal of the diagonal matrix $X$,
the frequency representation of this matrix is
\begin{eqnarray}
    \widehat{X}_{jj} &\equiv& \frac{1}{N^{2}}\mathcal{F}_{jl} X_{jj} \mathcal{F}_{jl}^{\dagger},
    \label{eq:Xjj}
\end{eqnarray}
where $\mathcal{F}_{jl}$ is the discrete Fourier transform matrix (of size $N\times N$) given by
\begin{eqnarray}
    \mathcal{F}_{jl}=\Big[e^{-2\pi i j l/N} \Big]_{j,l=0,...,N-1}.
    \label{eq:MatrixDFT}
\end{eqnarray}

Additionally, the mean $\mu_{\widehat{\epsilon}_F|\widehat{F}}$ and covariance $\Gamma_{\widehat{\epsilon}_F|\widehat{F}}$ of the conditional flux noise $\widehat{\epsilon}_F | \widehat{F}$ are
\begin{eqnarray}
    \mu_{\widehat{\epsilon}_F|\widehat{F}} &=&  \frac{\widehat{\sigma}_F^2}{\widehat{\sigma}_F^2 + \Sigma_{\phi}(\omega)}\widehat{F}(\omega), \\
    \Gamma_{\widehat{\epsilon}_F|\widehat{F}} &=& \left({\widehat{\sigma}_F^{-2} + \Sigma_{\phi}^{-1}}\right)^{-1},
\label{eq:Gamma_epsilon_given_F}
\end{eqnarray}
where $\Sigma_\phi$ is the variance function (a diagonal matrix) of the noiseless total quasar image fluxes defined in Equation~\ref{eq:Sigma_F_phi}, and $\widehat{\sigma}_F^2$ is the variance of the observational flux noise.

We can now express the probability of observing the set of centroid measurements $x(t)$ given that we have already observed the set of flux measurements $F(t)$.
Using Equation~\ref{eq:P_x_given_F}, the density of a
complex multivariate normal distribution
as defined in Appendix \ref{sec:cmvn} and equations \ref{eq:mu_G_given_F} and \ref{eq:Gamma_G_given_F} for the mean and covariance of $\widehat{G}|\widehat{F}$, we obtain the following expression for the log-probability
\begin{eqnarray}
\label{eq:log_P_centroid_given_flux}
    \ln{P(x|F, \tau, \alpha_1, \alpha_2)} = \ln{Z} + \ln P(\widehat{G}|\widehat{F}, \tau, \alpha_1, \alpha_2) \hspace{2em}\\ 
    = \ln{Z} -\ln (\det[\pi\Gamma_{\widehat{G}|\widehat{F}}]) - (\widehat{G}-\mu_{\widehat{G}|\widehat{F}})^\dagger\Gamma_{\widehat{G}|\widehat{F}}^{-1}(\widehat{G}-\mu_{\widehat{G}|\widehat{F}}). \nonumber
\end{eqnarray}
Note that for practical purposes, the calculation
of $\log\det$ should not be done by calculating the determinant,
as it involves the product of many numbers, some of which may be small.
For a practical calculation of $\log\det$, see Appendix~C in Paper~I.

\subsection{Full statistical model}
\label{sec:full_model}

By combining the results of Sections \ref{sec:stat_flux} and \ref{sec:stat_centroid_given_flux} (equations \ref{eq:log_P_flux} and \ref{eq:log_P_centroid_given_flux}) and using the law of conditional probability, we can write the log-likelihood of observing the specific set of total flux and total centroid measurements, $F(t)$ and $x(t)$, as a function of the unknown model parameters, $\tau$, $\alpha_1$ and $\alpha_2$, as
\begin{eqnarray}
\label{eq:full_model}
    \ln P(x, F | \tau, \alpha_i, x_i) = \hspace{10em} \\ \nonumber 
    \ln P(x | F, \tau, \alpha_i, x_i) + \ln P(F | \tau, \alpha_i).
\end{eqnarray}\\

Finally, if we are interested in detecting a lens system, we would like to perform hypothesis
testing against the null hypothesis,
and calculate the log likelihood difference between the null hypothesis ($H_{0}$) and alternative hypothesis ($H_{1}$)
(\citealt{Neyman+Pearson1933_HypothesisTesing}).
In our case, the null hypothesis parameters are $\tau=0$, and $\alpha_{2}=0$.
Therefore, our code calculates and minimizes
\begin{equation}
    \Delta\ln{\mathcal{L}} = -\ln{\mathcal{L}}(D|H_{1}) + \ln{\mathcal{L}}(D|H_{0}),
\end{equation}
where $D$ is the data $\{x(t), F(t)\}$.

\subsection{Extension to 2 dimensions}
\label{sec:2D}

Equation~\ref{eq:full_model} is written for the one dimensional case
(i.e., the lens and images are on the same line, whose direction is known).
Since the two dimensions are correlated, extending Equation~\ref{eq:full_model} to 2-D
does~not entail merely adding  the likelihood in the second dimension, but involves
the covariance between the two dimensions.
The 2-D formula will be derived in a future publication.
Currently, in order to solve the 2-D case, we suggest to calculate the
likelihood for all possible rotations under the assumption that all the images are on the same line.
This assumption is accurate for a two-images lens.
This scheme is implemented in our code and simulations.

\section{Implementation details}
\label{sec:Implimintation}

Many of the implementation details of the flux and astrometry method
are similar to those of the flux only method.
Therefore, we refer the reader to Paper~I for additional discussion on this matter.

One of the main caveats of our new algorithm is that the calculation
of $\mathcal{L}(\widehat{G}|\widehat{F})$
assumes
that different frequencies are independent.
As discussed in Paper~I, this assumption is not accurate.
For unevenly spaced time series, different frequencies are correlated,
and even in the evenly spaced case, correlations are induced by the leakage
of power outside of the observed band into the measured frequencies.
Furthermore, real time series are non-cyclic, while the Fast Fourier Transform (FFT)
implementation is cyclic.
This introduces a step function around the Nyquist frequency.

In Paper~I this problem was solved using two methods - one was heuristic and the other accurate.
The accurate method involves calculating the full covariance matrix of the model between
all the measurements (e.g., Eq.~\ref{eq:LL_TD}).
The heuristic approach involves applying an end-matching operation on the data prior to processing.
In the end-matching technique, we subtract a slope from the time series
such that the first and last point are of the same height (e.g., flux).
In Paper~I, we further discuss the limitations of the end-matching technique.

Writing the full covariance matrix of $\mathcal{L}(\widehat{G}|\widehat{F})$
is a work in progress.
Here, we extend the end-matching technique to astrometry.
The main problem is that after the end-matching is applied to the photometric
data, the center-of-light position is no longer consistent with the
photometry.
Therefore, our goal here is to modify the astrometric position in such a way that
it will be consistent with the photometric measurements.

The photometric end-matching is of the form:
\begin{equation}
    F'(t) = F(t) - \eta(t),
    \label{eq:FluxEndMatching}
\end{equation}
where $F'(t)$ is the photometric time series after end-matching, and
\begin{equation}
    \eta(t) = \beta_1 t + \beta_2.
    \label{eq:eta}
\end{equation}
Here, $\beta_1$ and $\beta_2$ are the coefficients of the end-matching linear function.
For a single image light curve (i.e., the source light curve),
the end-matching process for the source light curve can be written
as
\begin{equation}
    f'(t) \approx f(t) - \frac{\beta_1 t + \beta_2}{\alpha_{1}+\alpha_{2}},
    \label{eq:ftt}
\end{equation}
where $f'(t)$ is the source light curve after approximate end-matching.
The reason this is an approximation is that the images have different time delays.
However, assuming $\tau$ is small enough and that the quasar has a red power spectrum
(low variability amplitude on short time scales), this is a reasonable approximation,
and we test the validity of this approximation using simulations (see \S\ref{sec:simulations}).
In this case, the end-matching correction for the numerator
of Equation~\ref{eq:x} (i.e., $G$) is
\begin{equation}
    \zeta(t) = (\alpha_{1}x_{1}+\alpha_{2}x_{2})\frac{\beta_1 t + \beta_2}{\alpha_{1}+\alpha_{2}}.
    \label{eq:zeta}
\end{equation}
Next, since
\begin{equation}
    x(t) = \frac{G(t)}{F(t)},
    \label{eq:xt}
\end{equation}
we can write the end-matching corrected center-of-light as
\begin{equation}
    x'(t) = \frac{G(t) - \zeta(t)}{F(t) - \eta(t)}.
    \label{eq:xtt_def}
\end{equation}
Here, $x'(t)$ is the center-of-light position that is consistent
with the flux end-matching operation.
By rearranging Equation~\ref{eq:xtt_def} and using Equation~\ref{eq:xt},
we find
\begin{equation}
    x'(t) = \frac{x(t) - \zeta(t)/F(t)}{1 - \eta(t)/F(t)}.
    \label{eq:xtt}
\end{equation}

In our code, and in all the simulations we present,
we apply the end-matching to the flux,
and Equation~\ref{eq:xtt} to the center-of-light position.
We find that this procedure improves the results significantly.

\section{Code and optimization}
\label{sec:Code}

We implement our algorithm in Python and MATLAB.
In the Python code, available from Github\footnote{URL},
we provide a reference code
that includes all the major steps of the algorithm,
including simulating the light curves, and calculate the likelihoods given the observations and free parameters.

The MATLAB code is also available from Github\footnote{https://github.com/EranOfek/TimeDelay}
as part of the MATLAB Astronomy \& Astrophysics Toolbox (MAAT\footnote{https://webhome.weizmann.ac.il/home/eofek/matlab/index.html}, \citealt{Ofek2014_MAAT}).
In addition to light curve simulations and calculation of the likelihoods, the MATLAB code includes fitting functions.

The fitting procedure is based on the following approach.
For every trial time delay, we call a function that minimizes the likelihood for all the other parameters, and returns the best-fit parameters and likelihood.
The default minimizer is based on the Broyden–Fletcher–Goldfarb–Shanno algorithm.

\section{Simulations}
\label{sec:simulations}

In this section, we demonstrate the results derived in \S\ref{sec:method}
and \S\ref{sec:Implimintation} by numerically applying them to synthetic quasar flux and centroid data.
The parameters of our simulated light curves were chosen to be close to the relevant parameters of lensed quasars. In our simulations, we mainly used two sets of parameters listed in Table~\ref{tab:sim_params}.
The first set has $1''$ image separation and is mainly for demonstration purposes,
while the second set has $0.5''$ separation, which is supposed to mimic the typical lensed quasar.
We set the relative photometric errors to $0.03$. This is somewhat higher than the $0.01-0.02$ precision typically achieved by current ground-based sky surveys (e.g., \citealt{Padmanabhan+2008_SDSS_ImprovedPhotometricCalibration}; \citealt{Ofek+2012_photCalib}; \citealt{Schlafly+2012_PS1_PhotometricCalibration}; \citealt{Masci+2019_ZTF_Pipeline}). This was chosen in order to include a possible contribution from microlensing noise.
The astrometric errors were assumed to be at the level of $0.01''$, consistent with the precision achievable with current ground-based sky surveys in a single epoch (e.g., \citealt{Ofek2019_Astrometry_Code}).

Table~\ref{tab:sim_params} lists the light curves' parameters used in the main simulations
presented in this paper.
\begin{table}
	\centering
	\caption{Each simulation is coupled with its null hypothesis ($H_{0}$) simulation.
In the $H_{0}$ simulation, we assume $\tau=0$, and $\alpha_{2}=0$.
Time span is the time span of the evenly spaced simulations,
and in all cases, we used a sampling of 1\,day.
$\sigma_{F}/F$ is the relative flux error.
Oversampling is the oversampling factor in the frequency of the simulated light curves.
The unevenly spaced simulations are discussed in \S\ref{sec:UnevenlySpaced}.}
	\label{tab:sim}
	\begin{tabular}{lcrr} 
	    \hline
	    Parameter & Units & Sim 1 & Sim 2 \\
	    \hline
	    $\tau$               & days    & 30     & 30 \\
$\alpha_{0}$         &         & 0      & 0  \\
$\alpha_{1}$         &         & 1      & 1   \\
$\alpha_{2}$         &         & 0.5    & 0.5 \\
$x_0$                & $''$    & 0      & 0   \\
$x_1$                & $''$    & 0.2    & 0.1  \\
$x_2$                & $''$    & $-0.8$ & $-0.4$ \\
$y_{i}$              & $''$    & 0      & 0     \\
$\gamma$             &         & 2      & 2     \\
$\sigma_{F}/\langle F \rangle$    &  & 0.03 & 0.03 \\
$\sigma_{x}$         & $''$                & 0.01 & 0.01 \\
Time span ($n_{\rm t}$)      & days   & 1000    & 300 \\
Oversamlping         &        & 10      & 10 \\
std$(F)/\langle F \rangle$ &  & 0.1-0.15 & 0.1-0.15\\
		\hline
	\end{tabular}
\end{table}

The light curves are generated using the prescription in Appendix~\ref{sec:GeneratingSimulations}.
The generation of unevenly spaced light curves is further discussed in \S\ref{sec:UnevenlySpaced}.
In all the simulations, we apply the end-matching process (see \S\ref{sec:Implimintation}) to the data
prior to the analysis.
For the calculation of $\mathcal{L}(F)$
we use the Fourier space formula (Eq.~\ref{eq:logP_F_par}).

In \S\ref{sec:Example} we present
an example for a single simulated light curve and center-of-light curve,
and demonstrate our method on this light curve.
In \S\ref{sec:FittingUnknownRotation}
we demonstrate that our method can identify a lensed quasar for which
the images' on-sky 2-D positions
and fluxes are unknown.
In \S\ref{sec:FalseAlarms} we simulate
the usability of our method
as a lensed quasar detector.
In \S\ref{sec:WrongGamma} we investigate
the sensitivity of our method
when $\gamma$ is unknown,
while in \S\ref{sec:LargePhotErrors}
we discuss its sensitivity to
other parameters,
and in \S\ref{sec:UnevenlySpaced} we show some preliminary results from
applying this method to unevenly spaced time series.

\subsection{Example for simulated light curve and center-of-light position}
\label{sec:Example}

Here we provide an example for a simulated light and center-of-light curves as a function of time.
We also show
that our method converges to the correct solution.
We are using the parameters 
for simulation number 1
listed in Table~\ref{tab:sim_params}, to simulate an evenly spaced time series.
This simulation has a large image separation ($1''$) and hence
it is easier to visualize.

\begin{figure}
\centerline{\includegraphics[width=8cm]{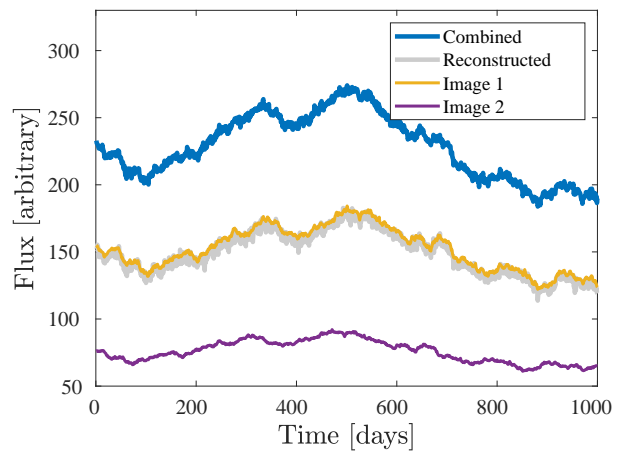}}
\caption{Simulated light curves of lensed quasar image number 1 (yellow), image number 2 (purple), and the noisy combined light curve (blue).
The gray line shows the reconstructed light curve of
the source (i.e., image number 1) based on Equation~\ref{eq:TemporalRepresentation}.
\label{fig:Ast_Sim1_Example_Ft}}
\end{figure}

\begin{figure}
\centerline{\includegraphics[width=8cm]{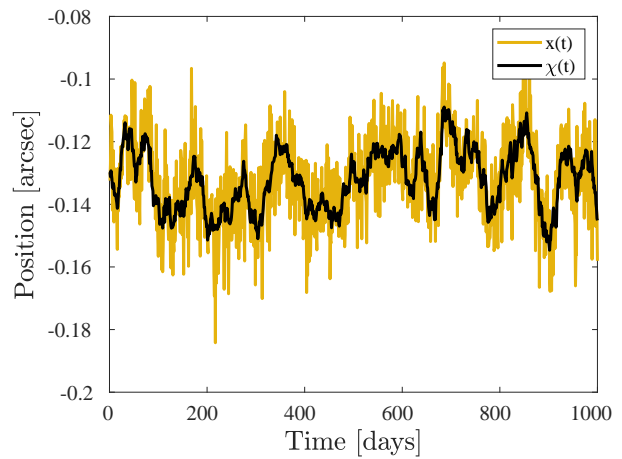}}
\caption{The noiseless center-of-light position (black),
and the noisy one (orange), plotted for
the same simulation as in Figure~\ref{fig:Ast_Sim1_Example_Ft}.
\label{fig:Ast_Sim1_Example_xt}}
\end{figure}

In Figure~\ref{fig:Ast_Sim1_Example_Ft} we present
a simulated light curve (no end-matching),
for the two images (noiseless) and combined light (with noise).
The gray line shows the reconstructed $f_{\rm rec}(t)$
light curve, based on Equation~\ref{eq:TemporalRepresentation}.
Figure~\ref{fig:Ast_Sim1_Example_xt},
shows the $\chi(t)$ and $x(t)$ (without end-matching),
of the same simulation.

\begin{figure}
\centerline{\includegraphics[width=8cm]{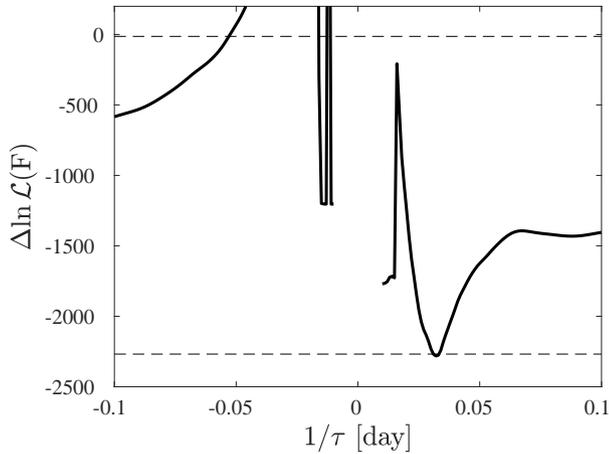}}
\caption{$\Delta\ln\mathcal{L}(F|\alpha_{i}, x_{i}, \theta=0 )$ as a function
of $1/\tau$ for the same simulation as in Figure~\ref{fig:Ast_Sim1_Example_Ft}.
The upper horizontal dashed line shows the theoretical $3\sigma$ confidence
threshold for detection assuming a $\chi^{2}/2$ distribution with nine degrees of freedom.
The lower horizontal dashed line is the $1\sigma$ confidence threshold above
the minimum of $\Delta\ln\mathcal{L}$ assuming a $\chi^{2}/2$ distribution with nine degrees of freedom.
The plot suggests that the time delay is recovered with high confidence and high accuracy.
\label{fig:Ast_Sim1_Example_DL_Tau}}
\end{figure}

\begin{figure}
\centerline{\includegraphics[width=8cm]{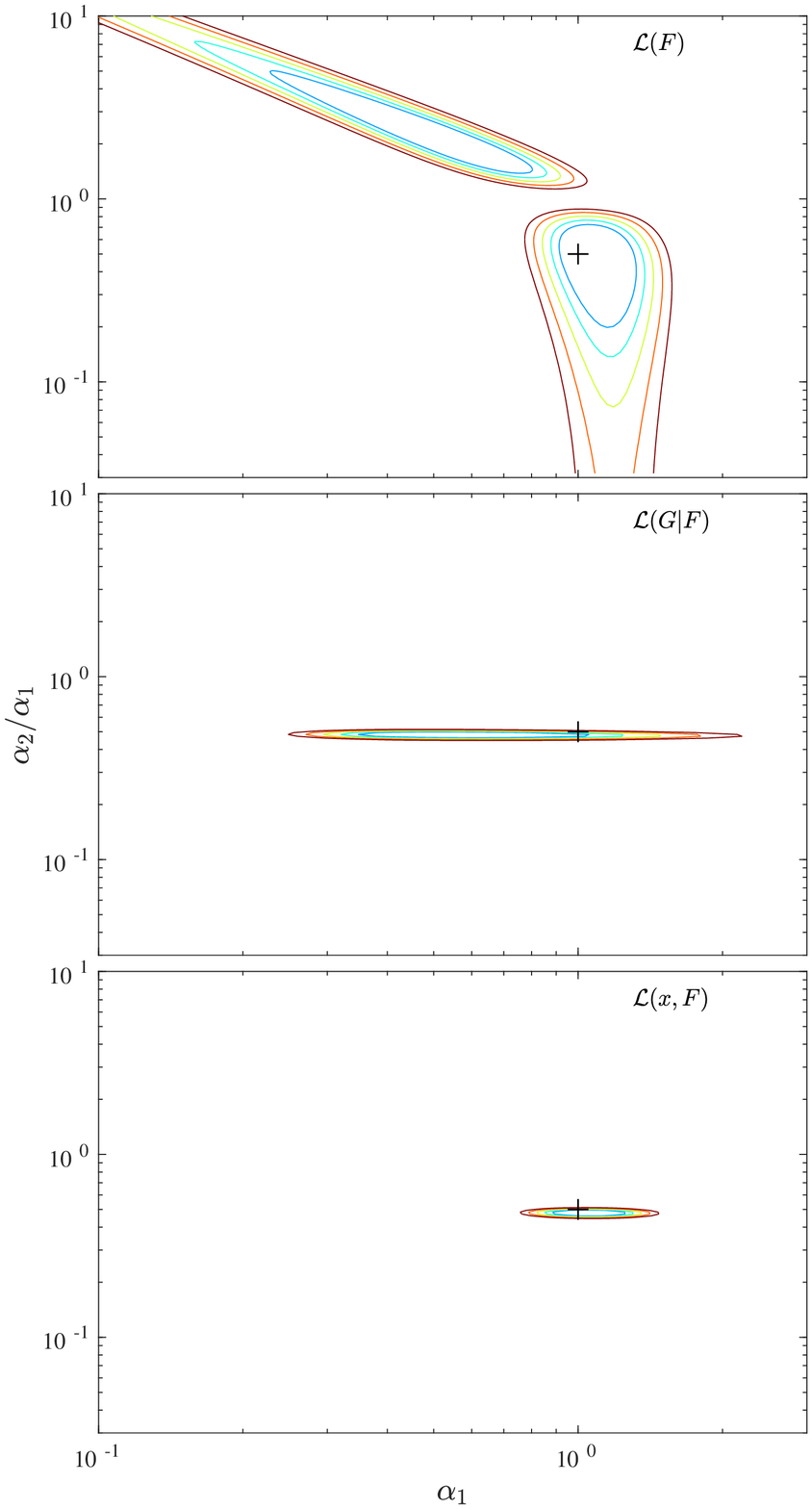}}
\caption{Contours of $\Delta\ln\mathcal{L}$ as a function of $\alpha_{1}$ and $\alpha_{2}/\alpha_{1}$,
when the other parameters are fixed to their true value,
for the same simulation as in Figure~\ref{fig:Ast_Sim1_Example_Ft}.
The upper plot shows the $\Delta\ln\mathcal{L}(F)$ (see also Paper~I),
the middle plot shows the $\Delta\ln\mathcal{L}(G|F)$,
and the lower plot shows the combined likelihood $\Delta\ln\mathcal{L}(x,F)$.
The contours show the 1,2,3,4,5-$\sigma$ confidence levels, assuming
a $\chi^{2}/2$ distribution with nine degrees of freedom.
The nine degrees of freedom were chosen in order to mimic the number of unknown
parameters (i.e., $\tau$, $\alpha_{0}$, $\alpha_{1}$, $\alpha_{2}$, $x_0$, $x_1$, $x_2$, $\theta$, $\gamma$) 
The plus marker shows the true value of the simulated $\alpha_{1}$ and $\alpha_{2}/\alpha_{1}$.
\label{fig:Ast_Sim1_A1A2_FGx}}
\end{figure}

For the same simulation, we present in Figure~\ref{fig:Ast_Sim1_Example_DL_Tau} the
$\Delta\ln\mathcal{L}$ as a function of $\tau$,
where in the $H_{1}$ fitting, for each $\tau$, we kept $\alpha_{i}$, and $x_{i}$
as free parameters in the fit,
and we assumed that $\gamma$ and
the on-sky rotation angle ($\theta$) of the lensed
images are known.
Here, we apply the end-matching procedure (see \S\ref{sec:Implimintation}), as otherwise,
the probability of success will be significantly lower.
Unless specified otherwise, the likelihood
as a function of $\tau$ is calculated
from $-0.1$ to $-0.01$\,day$^{-1}$,
and from $0.01$ to $0.1$\,day$^{-1}$,
in steps of $1/n_{\rm t}$\,day$^{-1}$, where $n_{\rm t}$
is the number of data points.
For the same simulated light curve, we show in Figure~\ref{fig:Ast_Sim1_A1A2_FGx}
the $\Delta\ln\mathcal{L}$ contours as a function
of $\alpha_{1}$ and $\alpha_{2}$,
while fixing $\alpha_{0}$ and $x_{i}$ to their true value.
The upper panel shows the likelihood contours of $\Delta\ln\mathcal{L}(F|\alpha_{1},\alpha_{2})$
(i.e., the flux-only likelihood; see also Paper~I).
The middle panel is for $\Delta\ln\mathcal{L}(G|F,\alpha_{1},\alpha_{2})$,
while the lower panel shows
the combined likelihood
$\Delta\ln\mathcal{L}(x,F|\alpha_{1},\alpha_{2})$.
These plots suggest that the flux plus astrometry method is considerably
more powerful than the flux-only method presented in Paper~I.

\subsection{Fitting when the rotation is unknown}
\label{sec:FittingUnknownRotation}

Here we test the ability of our method to identify a lensed quasar
when the position angle on the sky of the two brightest images,
and the position and fluxes of the images and lensing galaxy
are unknown.

\begin{figure}
\centerline{\includegraphics[width=8cm]{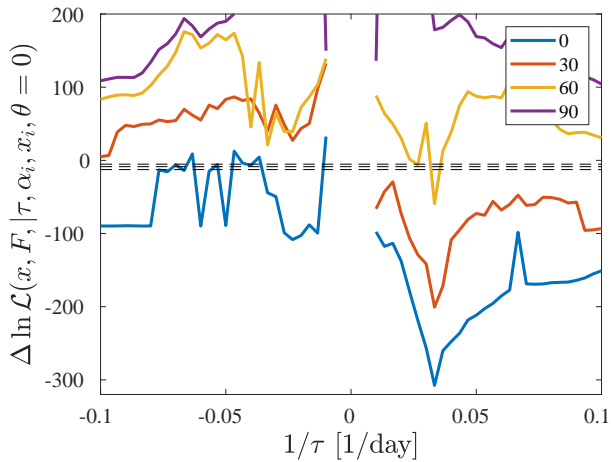}}
\caption{The $\Delta\ln\mathcal{L}$ as a function of $1/\tau$,
for a single simulated time series, after projecting $x(t)$
using rotation angles ($\theta$) of 0, 30, 60 and 90 degrees.
The dashed lines shows the theoretical threshold for detection based on
$\chi^{2}/2$ distribution with nine degrees of freedom.
\label{fig:Ast_Sim2_DL_TauRot_T300}}
\end{figure}

\begin{figure}
\centerline{\includegraphics[width=8cm]{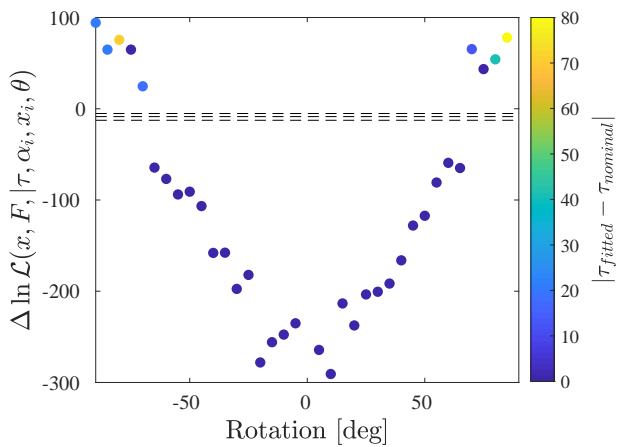}}
\caption{The minimum $\Delta\ln\mathcal{L}$, over the inspected $\tau$
as a function of the position angle ($\theta$); color coding
shows the absolute difference of the best fit $\tau$ from its nominal value.
The dashed lines are like in Figure~\ref{fig:Ast_Sim2_DL_TauRot_T300}.
\label{fig:Ast_Sim2_DL_Rot_T300}}
\end{figure}

We generate a simulated light curve using the parameters for simulation number 2
(Table~\ref{tab:sim_params}).
For every position angle ($\theta$) in the range of $-90$\,deg to 90\,deg,
with steps of 5\,deg, we rotate $x(t)$ and $y(t)$
by this position angle and use the projected $x(t)$.
Next, given the combined light curve and the projected $x(t)$,
we apply end-matching to $F(t)$ and the projected $x(t)$
(see \S\ref{sec:Implimintation}), and then
calculate the $\Delta\ln\mathcal{L}$ as a function of $1/\tau$.
As before, for each $\tau$, we keep $\alpha_{i}$ and $x_{i}$ as free parameters.
Figure~\ref{fig:Ast_Sim2_DL_TauRot_T300} shows the $\Delta\ln\mathcal{L}$ as a function of $\tau$,
for position angles of 0, 30, 60 and 90\,deg.
Figure~\ref{fig:Ast_Sim2_DL_Rot_T300} presents the minimum $\Delta\ln\mathcal{L}$
as a function of the position angle, where the color coding
shows the absolute difference of the best fit $\tau$ from its nominal value.
These figures demonstrate that our method can recover the correct
position angle and it works well even when the position angle is erroneous
by about 30\,deg.
Furthermore, this means that our method can potentially work
for systems with image separations below $0.1''$.

\subsection{False alarms}
\label{sec:FalseAlarms}

\begin{figure}
\centerline{\includegraphics[width=8cm]{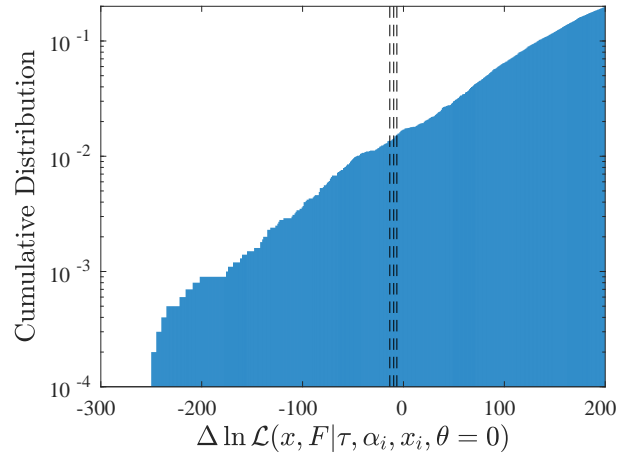}}
\caption{The cumulative histogram of the minimum
value of $\Delta\ln\mathcal{L}$ as a function of $1/\tau$ of each $H_0$ simulation.
The vertical dashed lines show the theoretical threshold for detection based on
a $\chi^{2}/2$ distribution with nine degrees of freedom.
\label{fig:Ast_Sim2_h0}}
\end{figure}

To test the usability of the new method as a detector,
here we attempt fitting the time delay to light curves and position curves
of a single quasar (i.e., the null hypothesis).
To do so, we generate $10^{4}$ simulations based on the
parameters for simulation number 2 (Table~\ref{tab:sim_params}),
but with $\tau=0$ and $\alpha_{2}=0$.
Figure~\ref{fig:Ast_Sim2_h0} presents the cumulative histogram of the minimum
value of $\Delta\ln\mathcal{L}$ as a function of $1/\tau$ of each simulation.

It is clear that in most cases the $\Delta\ln\mathcal{L}$ is positive,
as it should be,
meaning that our method preferred the null hypothesis over the alternative hypothesis.
We also see that the cumulative probability of $\Delta\ln\mathcal{L}$
does~not follow exactly the expectation from the $\chi^{2}/2$ distribution.
This is expected, for three reasons:
First, the hypotheses are not nested (see Paper~I).
Second, we here use the end-matching approximation
instead of the full covariance matrix.
Third, each simulation includes about 180 trials.
However, these trials are correlated
and the effective number of independent trials is lower.
Nevertheless, we see that the $\Delta\ln\mathcal{L}$
gives an approximate indication that the null hypothesis is rejected,
and that only $\sim1$\% of the simulations resulted in a negative $\Delta\ln\mathcal{L}$.
An important conclusion is that it is recommend to apply this method
along with simulations in order to calibrate the $\Delta\ln\mathcal{L}$.
This problem will become more important when dealing with unevenly spaced time series (\S\ref{sec:UnevenlySpaced}).

\subsection{Sensitivity to the wrong $\gamma$}
\label{sec:WrongGamma}

So far, our simulations assume that $\gamma$ is known.
However, quasar power spectra show large diversity with
a wide range of power-law indices in the range of 1.5 to 3.5
(e.g., \citealt{Mushotzky+2011_AGN_PowerSpectra_KeplerLC};
\citealt{Smith+2018_AGN_KeplerLC_PowerSpectrum}).
In principle, $\gamma$ can be fitted (see also Paper~I).
However, it is possible that the quasar power spectra
are not well described by a single power-law.
Therefore, we seek to verify that our method
has the potential to work even if the shape of quasar power spectra
is not exactly known.

\begin{figure}
\centerline{\includegraphics[width=8cm]{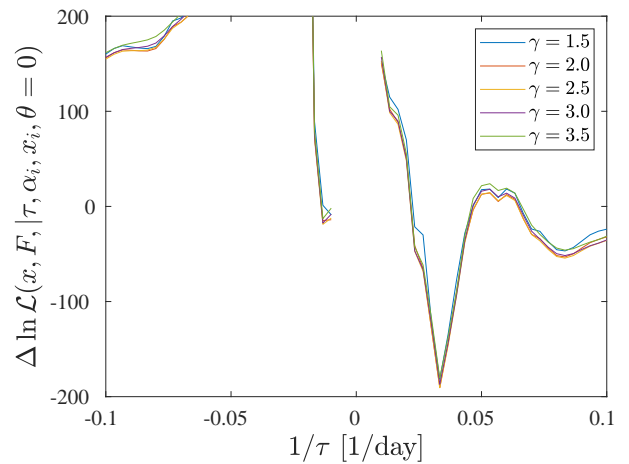}}
\caption{The $\Delta\ln\mathcal{L}$ as a function of $1/\tau$. The light curves were simulated
using $\gamma=2$, but each line represents a fit that assumes a different value of $\gamma$ (as indicated in the legend).
\label{fig:Ast_Sim2_Tau_DLL_wronggamma}}
\end{figure}

Figure~\ref{fig:Ast_Sim2_Tau_DLL_wronggamma} shows the $\Delta\ln\mathcal{L}$ as a function of $1/\tau$
using the parameters of simulation 2 in Table~\ref{tab:sim_params}.
However, here we attempted fitting the model while assuming different values
of $\gamma$.
We see that the fitting is not very sensitive to the assumed $\gamma$,
and our method can be used as a detector
and time-delay estimator
even if we do~not have accurate knowledge
of the light curve statistical model.

\subsection{Sensitivity to other parameters}
\label{sec:LargePhotErrors}

\begin{figure}
\centerline{\includegraphics[width=8cm]{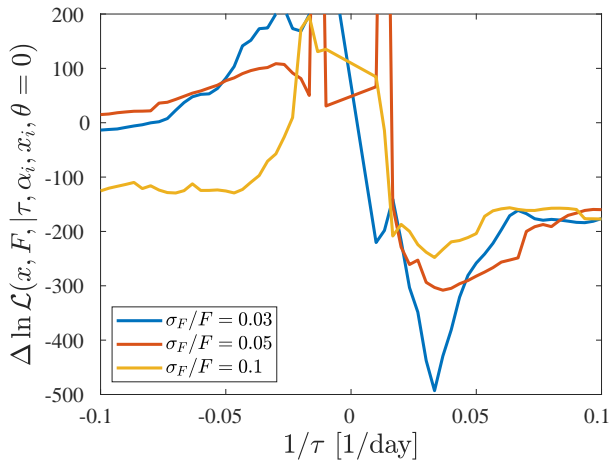}}
\caption{The $\Delta\ln\mathcal{L}$ as a function of $1/\tau$, for three simulations,
each one with a different photometric noise level (as indicated in the legend).
\label{fig:Ast_Sim2_sensSigmaF}}
\end{figure}

Here we present some of the tests we conducted in order to evaluate the sensitivity of our method to some of the other parameters
in the problem.
Our tests show that the method can work even if the flux ratio
is low (e.g., $\alpha_{2}/\alpha_{1}\gtorder0.03$),
or the photometric noise is high (e.g., $\sigma_{\rm F}/F\ltorder0.1$).
For example, in Figure~\ref{fig:Ast_Sim2_sensSigmaF}, we present
the $\Delta\ln\mathcal{L}$ as a function of $1/\tau$ for three
simulations based on the parameters of simulation number 2 (Table~\ref{tab:sim_params}),
but with different $\sigma_{\rm F}/F$ values.

It is clear that our method can potentially work even in cases of very noisy
photometric data.
This may be relevant to systems that are strongly affected by microlensing variability.
We note, however, that microlensing variability has
several time scales like caustic crossing (e.g., 3\,months) and Einstein radius
crossing (e.g., 10\,yr; \citealt{Wambsganss2001_LensedQuasars_Microlensing}).
Therefore, microlensing noise is supposed to be correlated, while our simulations
assumed white noise.

\subsection{Unevenly spaced time series}
\label{sec:UnevenlySpaced}

So far, we tested our method on evenly spaced data.
Here, we present some first attempts to apply this algorithm to unevenly
spaced measurements.
Unlike the method presented in Paper~I, we do~not yet have a full covariance
formalism for our flux plus astrometry method.
Therefore, in order to test our method on unevenly spaced data
we were forced to interpolate our light and position curves
into an evenly spaced grid.
This, of course, introduces some problems.
For example, we expect interpolated points to be correlated,
and this will affect the effective number of independent observations.

We start by generating an evenly spaced light curve with a sampling of $0.1$\,day.
Next, we generate an unevenly spaced grid, and interpolate our densely sampled light curve
onto the unevenly spaced grid.
The unevenly spaced grid is generated using the following approach:
We generate an evenly spaced grid with time differences of 1\,day, and add some
random Gaussian noise to the times with a standard deviation of $0.05$\,day.
Next, we divide the obtained times by the Lunar synodic period and remove points
whose phase is between 0.8 and 1.
We also remove points whose annual
phase is between 270 and 365 days.
We choose the total length of the light curve to be about two years.
Next, we resample the unevenly spaced light curves
into a uniform grid with 1\,day steps, and apply our method to the evenly spaced
light curves.

\begin{figure}
\centerline{\includegraphics[width=8cm]{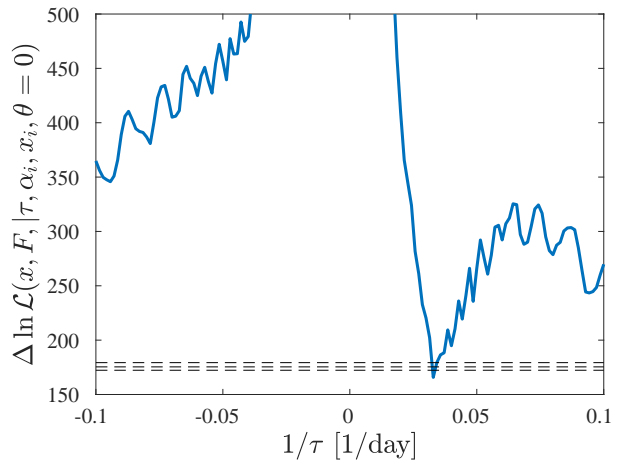}}
\caption{The mean of $\Delta\ln\mathcal{L}$ as a function of $1/\tau$, calculated over 100 unevenly spaced simulated light curves.
Each light curve spans two years.
The horizontal dashed lines show the minimum $\Delta\ln\mathcal{L}$ plus the 1,2,3-$\sigma$ confidence levels
on the best fit value, with
the confidence levels calculated assuming a $\chi^{2}/2$ distribution with nine degrees of freedom.
\label{fig:Ast_Sim_Unevenly}}
\end{figure}

In Figure~\ref{fig:Ast_Sim_Unevenly}, we present the mean of $\Delta\ln\mathcal{L}$ as a function of $1/\tau$, calculated over 100 simulated light curves,
where the parameters of simulation number 2 in Table~\ref{tab:sim_params} were used.
As before, $\alpha_{i}$ and $x_{i}$ were fitted, but we assumed the correct
rotation angle $\theta=0$.
The dashed lines mark the minimum in $\Delta\ln\mathcal{L}$ plus the
theoretical $\Delta\chi^{2}/2$ with nine degrees of freedom,
for one sided 1, 2, and 3$\sigma$.
We see that our method recovers the true period with satisfactory precision.
However, one feature of this plot is that the $\Delta\ln\mathcal{L}$ is positive.
This is presumably due to the interpolation, which introduces a correlation between
the points and effectively modifies the number of independent points in the light curve.
Therefore, in order to use this version of the method as a detector, one will probably need to rely
on simulations in order to calibrate the value of $\Delta\ln\mathcal{L}$.

\section{Discussion}
\label{sec:disc}

We present a novel method to identify lensed quasars and supernovae, and simultaneously measure their time delays, using
time series of the unresolved combined flux and astrometric position.
A subset of this method that uses only
the combined flux measurements is presented in Paper~I.
Our method utilizes the fact that, due to the time delay between lensed images, any flux variations in the source will induce
astrometric variations in the center-of-light position.
Unlike methods that rely on accurate knowledge of the PSF (e.g., forward modeling and deconvolution), this method may work
even if the image separation is well below
the instrument's resolution.
Our simulations suggest that this method even work when the image separation
is as low as $0.1''$,
assuming that the imaging data from which the center-of-light is measured
is Nyquist sampled\footnote{A Gaussian is not a band-limited function, and hence for a Gaussian PSF, the Nyqusit sampling is undefined. However, practically, for our applications sampling of a $\gtorder2$ pixels per full width at half the maximum of the PSF
is enough (see, e.g., \citealt{Ofek2019_Astrometry_Code}).}.

Applying these methods to data from previous, existing, and upcoming sky surveys (e.g., \citealt{Law+2009_PTF}; \citealt{Chambers+2016_PS1_Surveys}; \citealt{Bellm+2019_ZTF_Overview}; \citealt{Tyson+2001_LSST};
\citealt{Abell+2009_LSST_ScienceBook_Ver2};
\citealt{Tonry2011_ATLAS_SurveyCapability};
\citealt{Ofek+BenAmi2020_Grasp_SkySurvrys_CostEffectivness})
has the potential to uncover a significant fraction of all lensed quasars
and measure their time delays.
This method may also enable the efficient
discovery of lensed supernovae.

We derive the analytic formalism of this
method, and provide the formulae needed in order to apply it to real data.
We also provide Python and MATLAB reference implementations.
We test our methods on simulated data, and in future papers we will attempt to apply them to real data.
Our simulations suggest that applying these methods to real data will likely require
$\gtorder100$ photometric and astrometric measurements per source,
with precision better than about 10\% in photometry.

There are several caveats and extensions that require more work and analysis.
Among these are:
(i) quantifying the effect of microlensing variability on the accuracy of our methods:
The fact that in the presence of microlensing variability, and when the images are well resolved, quasar time delays are currently being successfully measured suggests that this is not a crucial point.
(ii)  The present version of our method requires applying the heuristic end-matching process to the data prior to the analysis.
This problem can be fixed by deriving our method, including the full covariance of the problem (like was done in Paper~I).
This point will also affect the treatment of unevenly spaced data.
(iii) The likelihood function should be extended to the full 2-D case. This will enable us to use this method to measure multiple time delays.

\section*{Acknowledgements}

We thank Oren Raz, Barak Zackay, Brad Cenko, and Ariel Goobar for many enlightening discussions.
E.O.O. is grateful for the support of
grants from the 
Willner Family Leadership Institute,
Ilan Gluzman (Secaucus NJ), Madame Olga Klein - Astrachan,
Minerva foundation,
Israel Science Foundation,
BSF, BSF-transformative, Israel Ministry of Science,
Weizmann-Yale, and Weizmann-UK.

\section*{Data Availability}

The scripts used to simulate and test the method are available via Github as indicated in the main text.

\bibliographystyle{mnras}
\bibliography{papers.bib}


\newpage
\onecolumn

\appendix

\section{Some properties of real and complex multivariate normal distributions}
\label{sec:cmvn}

\subsection{The multivariate normal distribution}

A real valued random vector $x \in \mathbb{R}^n$ is said to be distributed as a multivariate normal (MVN) if it has the following probability density function (PDF)

\begin{equation}
P(x|\mu_x, \Sigma_x) = \frac{e^{-\frac{1}{2}(x-\mu_x)^T\Sigma_x^{-1}(x-\mu_x)}}{\sqrt{\det[2\pi\Sigma_x]}},
\end{equation}
where $\det[M]$ denotes the determinant of matrix $M$ and the $T$ sign denotes a matrix transpose. The MVN distribution is characterized by the following mean vector and covariance matrix:
\begin{eqnarray}
\mu_x &\equiv& \operatorname{E}\left[x\right], \\
\Sigma_x &\equiv& \operatorname{E}\left[(x-\mu_x)(x-\mu_x)^T\right].
\end{eqnarray}
We denote the fact that the random vector $x$ has the above distribution by writing $x \sim N(\mu_x, \Sigma_x)$.

\subsection{The complex multivariate normal distribution}

A complex valued random vector $z \in \mathbb{C}^n$ is said to be distributed as a complex multivariate normal (CMVN) if the real valued random vector $(\mathrm{Re}(z), \mathrm{Im}(z)) \in \mathbb{R}^{2n}$ of its real and imaginary coefficients is distributed as a multivariate normal (MVN). In such a case, the distribution of $z$ can generally be characterized by the following complex mean vector $\mu_z$, real covariance matrix $\Gamma_z$ and \textit{relation} matrix $C_z$:
\begin{eqnarray}
\mu_z &\equiv& \operatorname{E}\left[z\right], \\
\Gamma_z &\equiv& \operatorname{E}\left[(z-\mu)(z-\mu)^\dagger\right], \label{eq:cov_cmvn}\\
C_z &\equiv& \operatorname{E}\left[(z-\mu)(z-\mu)^T\right], \label{eq:relation_matrix}
\end{eqnarray}
where the $\dagger$-symbol indicates the transpose complex-conjugate. When $C_z = 0$, the PDF of $z$ is the following:
\begin{equation}
P(z|\mu_z, \Gamma_z) = \frac{e^{-(z-\mu_z)^\dagger\Gamma_z^{-1}(z-\mu_z)}}{\det[\pi\Gamma_z]}.
\end{equation}
We denote the fact that the complex random vector $z$ has the above distribution by writing $z \sim CN(\mu_z, \Gamma_z)$.\\

\subsection{Affine transformation of MVN and CMVN random vectors}
\label{sec:affine_tranformation}

An MVN distribution $x \sim N(\mu_x, \Sigma_x)$ has the following affine transformation property:
\begin{eqnarray}
\label{eq:mvn_affine}
    x' &=& Ax + b \nonumber \\
    x' &\sim& N(b + A\mu_x, A\Sigma_x A^T).
\end{eqnarray}
A CMVN distribution $z \sim CN(\mu_z, \Gamma_z)$ has the following affine transformation property:
\begin{eqnarray}
\label{eq:cmvn_affine}
    z' &=& Az + b \nonumber \\
    z' &\sim& CN(b + A\mu_z, A\Gamma_z A^\dagger).
\end{eqnarray}
 Additionally, the relation matrix $C_z$ (Equation~\ref{eq:relation_matrix}) transforms as follows, $C_{z'} = A C_{z} A^T$.
 
 \subsection{Conditioning an MVN on some of its coordinates}
 
 If $x = (x_1, x_2)^T$ is an MVN column random vector in $\mathbb{R}^n$ composed of $x_1 \in \mathbb{R}^m$ and $x_2 \in \mathbb{R}^k$ and 
 
\begin{equation}
  \begin{pmatrix}
    x_1 \\
    x_2
  \end{pmatrix}
\sim N\left(
  \begin{pmatrix}
    \mu_1 \\
    \mu_2
  \end{pmatrix},
  \begin{pmatrix}
    \Sigma_{11} & \Sigma_{12} \\
    \Sigma_{21} & \Sigma_{22}
  \end{pmatrix}
\right),
 \end{equation}
where $\mu_1$ and $\mu_2$ are the mean vectors of $x_1$ and $x_2$ and $\Sigma_{11} \in \mathbb{R}^{m\times m}$, $\Sigma_{12} \in \mathbb{R}^{m\times k}$, $\Sigma_{21} \in \mathbb{R}^{k\times m}$ and $\Sigma_{22} \in \mathbb{R}^{k\times k}$ are the appropriate sub-matrices of $x$'s full covariance matrix, then the random vector $x_1|x_2$, which has the conditional distribution $P(x_1|x_2)$, distributes as follows \citep[see][sec. 9.3]{vonMises1964_MathTheoryProbabilityAndStatistics}:
\begin{equation}
\label{eq:mvn_cond}
    x_1|x_2 \sim N\left(
        \mu_1+\Sigma_{12}\Sigma_{22}^{-1}\left(x_2-\mu_2\right),
        \Sigma_{11}-\Sigma_{12}\Sigma_{22}^{-1}\Sigma_{21}.
    \right)
\end{equation}

\subsection{Conditioning a Gaussian on its sum with another Gaussian}
\label{sec:cond_gaussian_on_sum}

Given two real scalar Gaussian random variables, $x\sim N(\mu_x, \sigma_x^2)$ and $y\sim N(\mu_y, \sigma_y^2)$, that are statistically independent, and defining $z \equiv x + y$, then the conditional distribution $x|z$ is also a Gaussian random variable having the following mean and variance:
\begin{eqnarray}
    \operatorname{E}\left[x|z\right] &=& \mu_x + \frac{\sigma_x^2}{\sigma_x^2+\sigma_y^2}(z-\mu_x-\mu_y) \\
    \mathrm{Var}\left[x|z\right] &=& (\sigma_x^{-2}+\sigma_y^{-2})^{-1}
\end{eqnarray}
This can be shown by using Equation~\ref{eq:mvn_affine} to find the MVN resulting from the linear transformation $(x_1, x_2)^T = \bigl( \begin{smallmatrix}1 & 0\\ 1 & 1\end{smallmatrix}\bigr)(x, y)^T = (x, z)^T$ and then applying Equation~\ref{eq:mvn_cond} to obtain the mean and variance of $x_1|x_2$ or $x|z$.

\section{The distribution $P(\widehat{G}|\widehat{F}, \tau, \alpha_1, \alpha_2)$}
\label{sec:distrib_G_given_F}

We wish to find $P(\widehat{G}|\widehat{F}, \tau, \alpha_i, x_i)$. For brevity, we write the distribution of $\widehat{G}$ conditioned on the observed $\widehat{F}$ as $\widehat{G} | \widehat{F}$. We recall the last line of Equation~\ref{eq:G_approx2}:
\begin{equation}
\label{eq:G_approx2_recall}
    \widehat{G}(\omega) = 
    \widehat{A}\widehat{F} + \widehat{F}*\widehat{\epsilon}_x + \widehat{\chi}*\widehat{\epsilon}_F - \widehat{A}\widehat{\epsilon}_F,
\end{equation}
which holds to first order in the noise terms. First, we note that the only remaining stochastic terms in $\widehat{G} | \widehat{F}$ are $\widehat{\epsilon}_x|\widehat{F}$, $\widehat{\epsilon}_F| \widehat{F}$, and $\widehat{\chi}| \widehat{F}$. We will now find the distribution of the last three terms appearing in Equation~\ref{eq:G_approx2_recall}. We will show in the following that $\widehat{G}|\widehat{F}$ is a sum of CMVN random vectors (to first order in the noise terms) and is, therefore, also a CMVN.\\

The distribution of the conditional flux noise in the frequency domain, $\widehat{\epsilon}_F|\widehat{F}$, can be deduced as follows. Writing the Fourier transform of Equation~\ref{eq:F},
\begin{equation}
    \widehat{F} = \widehat{\phi} + \widehat{\epsilon}_F,
\end{equation}
we see that $\widehat{F}$ is a sum of two statistically independent random vectors with zero mean normally distributed independent per-frequency components with independent real and imaginary parts. As we show in \S\ref{sec:cond_gaussian_on_sum}, this leads to $\widehat{\epsilon}_F|\widehat{F}$ being a CMVN with the following mean and covariance:
\begin{eqnarray}
\mu_{\widehat{\epsilon}_F|\widehat{F}} &=&  \frac{\widehat{\sigma}_F^2}{\widehat{\sigma}_F^2 + \Sigma_{\phi}(\omega)}\widehat{F}(\omega), \\
\Gamma_{\widehat{\epsilon}_F|\widehat{F}} &=&  \mathrm{Var}\left[\mathrm{Re}(\widehat{\epsilon}_F(\omega))|\widehat{F}\right] + 
\mathrm{Var}\left[\mathrm{Im}(\widehat{\epsilon}_F(\omega))|\widehat{F}\right] = \left({\widehat{\sigma}_F^{-2} + \Sigma_{\phi}^{-1}(\omega)}\right)^{-1},\\
C_{\widehat{\epsilon}_F|\widehat{F}} &=& 0,
\end{eqnarray}
where $\Sigma_\phi(\omega)$ is the variance function of the noiseless total quasar image fluxes defined in Equation~\ref{eq:Sigma_F_phi}. To show that the relation matrix $C_{\widehat{\epsilon}_F|\widehat{F}} = 0$, we use the fact that all of its off-diagonal elements are zero due to the statistical independence of $\widehat{\epsilon}_F|\widehat{F}$ at different frequencies. For the on-diagonal terms, one can show for a complex scalar random variable $z=\mathrm{Re}(z)+i\,\mathrm{Im}(z)$ that $\operatorname{E}\left[(z-\operatorname{E}\left[z\right]\right)^2] = 0$ when the real and imaginary parts of $z$ are statistically independent and have equal variance, as is the case for the real and imaginary parts of $\widehat{\epsilon}_F|\widehat{F}$ at each frequency.\\

We now express the third and fourth terms of Equation~\ref{eq:G_approx2_recall}. First, we rewrite the third term as follows:
\begin{equation}
    \mathcal{F}\left[\chi\right] * \widehat{\epsilon}_F = 
    \mathcal{F}\left[\frac{\mathcal{F}^{-1}\left[\widehat{A}\widehat{\phi}\right]}{\phi}  \right] * \widehat{\epsilon}_F = 
    \mathcal{F}\left[\frac{\mathcal{F}^{-1}\left[\widehat{A}(\widehat{F}-\widehat{\epsilon}_F)\right]}{F-\epsilon_F}  \right] * \widehat{\epsilon}_F \simeq 
    \mathcal{F}\left[\frac{\mathcal{F}^{-1}\left[\widehat{A}\widehat{F}\right]}{F}  \right] * \widehat{\epsilon}_F \equiv \widehat{\xi}*\widehat{\epsilon}_F.
    \label{eq:third_term}
\end{equation}
Here the first and second equalities result from definitions \ref{eq:F} and \ref{eq:x} and Equation~\ref{eq:G_of_omega}, the third equality is a first-order noise approximation and the final one defines $\widehat{\xi}$. Terms (3+4) of Equation~\ref{eq:G_approx2_recall} can now be expressed as
\begin{equation}
    \widehat{\xi}*\widehat{\epsilon}_F - \widehat{A}\widehat{\epsilon}_F = 
    \mathcal{F}[\xi \epsilon_F] - \widehat{A}\widehat{\epsilon}_F = 
    (\mathcal{F}X\mathcal{F}^{\dagger} - \widehat{\mathbb{A}})\widehat{\epsilon}_F \equiv 
    (\widehat{X} - \widehat{\mathbb{A}})\widehat{\epsilon}_F.
\end{equation}
Here, $\widehat{\mathbb{A}}$ is a diagonal matrix, having the values of $\widehat{A}$ on the diagonal,
$X$ is a diagonal matrix with the values of $\xi$ along the diagonal, $\widehat{X}$ is the Fourier representation of matrix $X$ and $\mathcal{F}$ is the discrete Fourier transform matrix. Using the above expression and the affine transformation property of CMVNs (\S \ref{sec:affine_tranformation}), we find the mean and covariance contribution of terms (3+4) of Equation~\ref{eq:G_approx2_recall} to $\widehat{G}|\widehat{F}$:
\begin{eqnarray}
    \mu_{3+4} &\equiv&  (\widehat{X} - \widehat{\mathbb{A}})\mu_{\widehat{\epsilon}_F|\widehat{F}} \label{mu_3_plus_4}\\
    \Gamma_{3+4} &\equiv& (\widehat{X} - \widehat{\mathbb{A}})\Gamma_{\widehat{\epsilon}_F|\widehat{F}}(\widehat{X} - \widehat{\mathbb{A}})^{\dagger} \label{Gamma_3_plus_4}
\end{eqnarray}

The centroid noise $\widehat{\epsilon}_x$ is a linear (Fourier) transformation of the zero mean Gaussian noise vector $\epsilon_x$ and is assumed to be statistically independent of $\widehat{F}$. We thus show that $\widehat{\epsilon}_x|\widehat{F}$ distributes as a CMVN and that $\mu_{\widehat{\epsilon}_x|\widehat{F}} = 0$ and $C_{\widehat{\epsilon}_x|\widehat{F}} = 0$ for strictly positive frequencies. To find the conditional covariance of the second term of Equation~\ref{eq:G_approx2_recall}, we rewrite it as follows:
\begin{equation}
    \widehat{F}*\widehat{\epsilon}_x = \mathcal{F}[F(t)\epsilon_x(t)] = \mathcal{F}[F \epsilon_x],
\end{equation}
where $\mathrm{F}$ is a diagonal matrix with $F(t)$ along the diagonal. Using the definition of the CMVN covariance matrix (Equation~\ref{eq:cov_cmvn}) and the fact that $\widehat{F}*\widehat{\epsilon}_x$ has a zero mean, we find the covariance contribution of the second term of Equation~\ref{eq:G_approx2_recall} to $\widehat{G}|\widehat{F}$:
\begin{equation}
    \Gamma_{2} \equiv \operatorname{E}\left[(\mathcal{F}\mathrm{F}\epsilon_x)(\epsilon_x^\dagger\mathrm{F}^\dagger\mathcal{F}^\dagger)\right] = \sigma_x^2(\mathcal{F}\mathrm{F}^2\mathcal{F}^\dagger) = \sigma_x^2(\mathcal{F}F)(\mathcal{F}F)^\dagger = \sigma_x^2\widehat{F}\widehat{F}^\dagger.
    \label{eq:Gamma_2}
\end{equation}

Finally, we may now add the two contributions to the mean of $\widehat{G}|\widehat{F}$, from the first term of Equation~\ref{eq:G_approx2_recall}, $\widehat{A}\widehat{F}$, and from terms (3+4):
\begin{equation}
\label{eq:mu_G_given_F_deriv}
    \mu_{\widehat{G}|\widehat{F}} = \widehat{A}\widehat{F} + (\widehat{X} - \widehat{\mathbb{A}})\mu_{\widehat{\epsilon}_F|\widehat{F}},
\end{equation}
and add the two contributions to the covariance of $\widehat{G}|\widehat{F}$ from the two statistically independent CMVNs resulting from the second term of Equation~\ref{eq:G_approx2_recall} and terms (3+4):
\begin{equation}
\label{eq:Gamma_G_given_F_deriv}
    \Gamma_{\widehat{G}|\widehat{F}} = \Gamma_2 + \Gamma_{3+4} = \sigma_x^2\widehat{F}\widehat{F}^\dagger + (\widehat{X} - \widehat{\mathbb{A}})\Gamma_{\widehat{\epsilon}_F|\widehat{F}}(\widehat{X} - \widehat{\mathbb{A}})^{\dagger}.
\end{equation}

\section{Generalization to the multi-image case} \label{sec:multi_image}

When more than two images of the quasar are present, some of the formulae of \S\ref{sec:method} need to be generalized to accommodate this. We start by updating equations \ref{eq:F} and \ref{eq:x} of the time domain total observed flux and centroid. Assuming $n$ images are present at sky positions $x_i$, each with flux factors $\alpha_i$ and time-delays $\tau_i$ relative to image 1, then
\begin{eqnarray}
\label{eq:flux_multi}
F(t) &=& \phi(t) + \epsilon_F(t) = \alpha_0 + \sum_{i=1}^n \alpha_i f(t+\tau_i) + \epsilon_f(t), \\
x(t) &=& \vec{\chi}(t) + \vec{\epsilon}_x(t) = \frac{\alpha_0 \vec{x}_0 + \sum_{i=1}^n \alpha_i \vec{x}_i f(t+\tau_i)}{\alpha_0 + \sum_{i=1}^n \alpha_i} + \epsilon_x(t),
\end{eqnarray}
and Equation \ref{eq:F_of_omega} for the noiseless and noisy flux in the frequency domain becomes
\begin{eqnarray}
    \widehat{\phi}(\omega) &=& \alpha_0 \delta(\omega) + \left(\sum_{i=1}^n \alpha_i e^{i\omega\tau_i}\right) \widehat{f}(\omega), \\
    \widehat{F}(\omega) &=& \widehat{\phi}(\omega) + \widehat{\epsilon}(\omega).
\end{eqnarray}
This allows us to update the expression for the variance of the noiseless total flux, Equation~\ref{eq:Sigma_F_phi}, as follows:
\begin{eqnarray}
\label{eq:phi_var_multi}
    \operatorname{E}\left[\widehat{\phi}(\omega)\widehat{\phi}^*(\omega)\right] \equiv \Sigma_\phi(\omega) &=& 
    \operatorname{E}\left[
    \left(\sum_{i=1}^n \alpha_i e^{i\omega\tau_i}\right)
    \widehat{f}(\omega)\widehat{f}^*(\omega)
    \left(\sum_{i=1}^n \alpha_i e^{-i\omega\tau_i}\right)
    \right] \\ \nonumber
    &=& \frac{1}{|\omega|^\gamma}
    \left[
    \sum_{i=1}^n \alpha_i^2 +
    2\sum_{i>j}\alpha_i\alpha_j \cos(\omega(\tau_i-\tau_j))
    \right],
\end{eqnarray}
which is valid for non-zero frequencies. Redefining $\Sigma_F(\omega) \equiv \Sigma_\phi(\omega) + \widehat{\sigma}_F^2$, we can also update Equation~\ref{eq:log_P_flux} for the log-likelihood of observing the total flux given the model parameters:
\begin{eqnarray}
\label{eq:log_P_flux_multi}
    \log P(F | \tau_i, \alpha_i) = -\frac{1}{2}\log \det[2\pi\Sigma_{F}] - 
    \sum_{\omega}\frac{|\widehat{F}(\omega)|^2}{2\Sigma_F(\omega)}.
\end{eqnarray}
Redefining the following frequency domain operator (originally defined in Equation~\ref{eq:diagnostic}),
\begin{equation}
\label{eq:operator_A_def_multi}
    \widehat{A}(\omega; \tau_i, \alpha_i) \equiv \frac{\sum_{i=1}^n \alpha_i x_i e^{i\omega\tau_i}}{\sum_{i=1}^n \alpha_i e^{i\omega\tau_i}}, 
\end{equation}
then the derivations of $\mu_{\widehat{G}|\widehat{F}}$ and $\Gamma_{\widehat{G}|\widehat{F}}$ (equations \ref{eq:mu_G_given_F_deriv} and \ref{eq:Gamma_G_given_F_deriv}) and the following likelihood functions (originally equations \ref{eq:log_P_centroid_given_flux} and \ref{eq:full_model}) remain valid up-to a change of arguments
\begin{eqnarray}
\log Z^{-1} P(x|F, \tau_i, \alpha_i) &=& \log P(\widehat{G}|\widehat{F}, \tau_i, \alpha_i) = -\log |\pi\Gamma_{\widehat{G}|\widehat{F}}| - (\widehat{G}-\mu_{\widehat{G}|\widehat{F}})^\dagger\Gamma_{\widehat{G}|\widehat{F}}^{-1}(\widehat{G}-\mu_{\widehat{G}|\widehat{F}}), \\
\log P(x, F | \tau_i, \alpha_i) &=& \log P(x | F, \tau_i, \alpha_i) + \log P(F | \tau_i, \alpha_i).
\end{eqnarray}

\section{Generating the simulated light and position curves}
\label{sec:GeneratingSimulations}

Our simulated light curves and center-of-light curves are generated using the following prescription:

\begin{itemize}

\item First, a frequency realization of the original quasar flux, $\widehat{f}(\omega)$, was pseudo-randomly drawn to have a red power spectrum at a set of discrete angular frequencies, $\omega \in (-\frac{n_{\rm t}}{2}+1, ..., 0, \frac{n_{\rm t} }{2}-1)\frac{2\pi}{T}$, where $n_{\rm t} = T/\Delta t$. Here, $T$ is the total time of the simulation, $\Delta t$ is the time interval between consecutive observations, and $n_{\rm t}$ is the number of temporal samples. To produce a red power spectrum, each frequency component's real and imaginary parts were randomly drawn to have a uniform phase around the full unit circle and a normally distributed magnitude with a mean zero and a standard deviation of $|\omega|^{-\gamma/2}$. Additionally, the constant flux level $n_{\rm t} f_{DC}$ was added to $\widehat{f}(0)$, where the pre-factor $n_{\rm t}$ is due to the definition of the discrete Fourier transform (DFT) used. To ensure the resulting temporal realization $f(t)$ has no imaginary components, negative frequency components of $\widehat{f}(\omega)$ were set to be complex conjugate symmetric to their matching positive components.

\item The resulting frequency domain noiseless total flux  is set (according to Equation~\ref{eq:Phi_F_of_omega}) to
$\widehat{\phi}(\omega) = (\alpha_1+\alpha_2 e^{i\omega\tau})\widehat{f}(\omega)$ and, additionally, the constant lensing galaxy flux $n_{\rm t} \alpha_0$ is added to the zero frequency component $\widehat{\phi}(0)$.

\item In a similar fashion to the generation of the red noise, the Gaussian noise $\widehat{\epsilon}_F(\omega)$ is generated in the frequency domain such that it would produce (following an inverse DFT) a real temporal flux noise $\epsilon_F(t)$ with a standard deviation $\sigma_F$. Adding the observational noise to $\widehat{\phi}(\omega)$ produced $\widehat{F}(\omega)$.

\item The $n_{\rm t}$ temporal samples of $f(t)$, $f_1(t)$, $f_2(t)$, $\phi(t)$, and $F(t)$ are computed (using an inverse DFT) from their frequency realizations $\widehat{f}(\omega)$, $\alpha_1\widehat{f}(\omega)$, $\alpha_2 e^{i\omega\tau}\widehat{f}(\omega)$, $\widehat{\phi}(\omega)$, and $\widehat{F}(\omega)$, respectively.
    
\item Working in a coordinate system where $x_0 = 0$, the noiseless total centroid is then computed in the temporal domain, $\chi(t) = \left(\alpha_1 f_1(t) + \alpha_2 f_2(t)\right) / \phi(t)$.

\item Real zero mean Gaussian noise $\epsilon_x(t)$ with a standard deviation of $\sigma_x$ is then added to produce the temporal observed total centroid $x(t) = \chi(t) + \epsilon_x(t)$.

\end{itemize}

This prescription generates light curves with cyclic boundary conditions, which is unrealistic.
In order to generate a non-cyclic simulation, we generate a light curve that is twice as long
as the desired light curve, and trim it to the needed duration.
In addition, by default, we generate light curves which are over sampled by a factor of
ten (see also Paper~I).
Next, after the light curve is generated, we check that it is positive in the entire period.
If not, we declare the light curve as invalid, and generate a new simulation, until a good light curve is generated.
This point is related to the unanswered question - is the power spectrum of quasar light curves
are power law in flux space or in log-flux space?
%
Finally, we add an option into our algorithm to select light curves
with specific standard deviation of flux divided by mean flux.

This algorithm generates evenly spaced data.
To generate unevenly spaced data, we simply generate a denser grid,
and interpolate it to the desired unevenly spaced times.
For light curves for which the power at long frequencies is small
(e.g., quasars), this approximation is reasonable.

\bsp	
\label{lastpage}

\end{document}